\documentclass{sf2a-conf2012}
\usepackage{graphicx}
\usepackage{hyperref}
\usepackage[]{natbib}  
\usepackage[cyr]{aeguill}
\usepackage{epstopdf}

\def\BibTeX{{\rm B\kern-.05em{\sc i\kern-.025em b}\kern-.08em
    T\kern-.1667em\lower.7ex\hbox{E}\kern-.125emX}}
\bibpunct{(}{)}{;}{a}{}{,}  

\begin{document}

\TitreGlobal{SF2A 2012}


\title{Twisted magnetic structures emerging from buoyancy instabilities}

\runningtitle{Twisted flux ropes and buoyancy instabilities}

\author{L. Jouve}\address{Universit\'e de Toulouse, UPS-OMP, CNRS, Institut de Recherche en Astrophysique et Plan\'etologie, Toulouse, France}

\author{L.J. Silvers}\address{Department for Mathematical Science, City University London, Northampton Square, London, EC1V 0HB, UK}

\author{M.R.E. Proctor}\address{Department of Applied Mathematics and Theoretical Physics, University of Cambridge,
Centre for Mathematical Sciences, Wilberforce Road, Cambridge CB3 0WA, UK}

\setcounter{page}{237}


\maketitle


\begin{abstract}
We here report calculations of magnetic buoyancy instabilities of a sheared magnetic layer where two separate zones are unstable. The idea is to study the possible generation of large-scale helical structures which could then rise through a stellar convection zone and emerge at the surface to create active regions. The calculations shown here are a follow-up of the work of \citet{Favier12} where the instability developed in a weakly magnetized atmosphere, consisting of a uniform field oriented in a different direction from the unstable layer below. Here, the top layer representing the atmosphere is itself unstable to buoyancy instabilities and thus quickly creates a more complex magnetic configuration with which the layer below will interact. We also find in this case that the accumulation of magnetic tension between the two unstable layers favors the creation of large-scale helical structures. 
\end{abstract}


\section{Introduction}

The idea of this work is to study the onset and development of magnetic buoyancy instabilities of a sheet of field in a configuration where large scale helical magnetic structures may emerge. Detailed calculations were performed by \citet{Favier12}, in which the buoyancy instability develops in a weakly magnetized atmosphere where the direction of the field varies with respect to the direction of the strong unstable field below. The idea was to understand how coherent structures could be created at the base of a stellar convection zone where a large-scale shear produces a strong toroidal magnetic field associated with a small poloidal component. The authors showed that from small-scale interchange instability, large-scale helical coherent magnetic structures could be created, thus giving a possible explanation for the coherent rise of magnetic flux from the base of the convection zone to the surface where bipolar active regions emerge.

Here, we investigate the case where the atmosphere above is itself unstable to buoyancy instabilities and naturally creates a more complex field configuration. The idea here is to study the possibility of the spontaneous creation of similar coherent structures when the atmospheric field is less idealized and organized at smaller scales. The calculation thus starts with an initial magnetic layer oriented in the $x$-direction, complemented with a transverse field $B_y$ such that two zones are initially made unstable to 2D and 3D perturbations.

\section{Model set up}

\subsection{Governing equations}

The evolution is governed by the compressible MHD equations, with heat transfer treated in the diffusion approximation. The polytropic atmosphere is considered to be convectively stable so as to focus on the instabilities driven by the magnetic field. Those equations are solved in a Cartesian domain in the same way as \citet{Matthews95}. In particular, the following dimensionless parameters are used:

$$
\sigma=\frac{\mu C_p}{K} \,\,\,\, , \,\,\,\, \zeta=\frac{\eta \rho_0 C_p}{K} \,\,\,\, , \,\,\,\, \theta=\Delta d/T_0 \,\,\,\, , \,\,\,\, m=\frac{g}{R\Delta}-1 \,\,\,\, , \,\,\,\, F=\frac{B_0^2}{p_0\mu_0}   \,\,\,\, , \,\,\,\, C_k=\frac{K}{\rho_0C_p(RT_0)^{1/2}}
$$

\noindent where $p_0$, $\rho_0$ and $T_0$ being the pressure, density and temperature at $z=0$, $\mu_0$ the magnetic permeability, $C_p$ the specific heat at constant pressure, $R$ the gas constant, $g$ the gravitational acceleration, $\Delta$ is the temperature gradient and $d$ is the depth of the box. The thermal conductivity $K$, viscosity $\nu$ and magnetic diffusivity $\eta$ are all assumed to be constant.

The equations then read:

\begin{equation}
\frac{\partial \rho}{\partial t}=-\nabla \cdot (\rho \bf u)
\end{equation}

\begin{equation}
\frac{\partial \rho {\bf u}}{\partial t}=-{\bf\nabla} (p+FB^2/2)+{\bf \nabla} \cdot (F {\bf BB} - \rho {\bf uu} + \sigma C_k {\bf {\tau}}) + (m+1) \theta \rho {\bf z}
\end{equation}

\begin{equation}
\frac{\partial {\bf B}}{\partial t}= {\bf \nabla \times} ( {\bf u \times B} - \zeta C_k {\bf \nabla \times B})
\end{equation}

\begin{equation}
\frac{\partial T}{\partial t}=-{\bf u \cdot \nabla}T - (\gamma - 1) T {\bf \nabla \cdot u} + \frac{\gamma C_k}{\rho} \nabla^2 T + \frac{C_k (\gamma -1)}{\rho}(\sigma \tau^2/2+F\zeta J^2)
\end{equation}

where $\bf \tau$ is the stress tensor and where the ideal gas law is used. The boundary conditions used are impermeable and stress-free at top and bottom, vertical magnetic field at top and bottom and constant flux at the bottom for the temperature.

\subsection{Initial field configuration}

The initial magnetic field configuration consists in a layer of magnetic field in which two spatially separated zones will be unstable with respect to buoyancy instabilities. The magnetic layer is located between $z=0.6$ and $z=0.8$ (the top of the domain being located at z=0). As shown by \citet{Newcomb61}, the criteria for 2D and 3D instabilities respectively read:

\begin{equation}
\rho \, \, \vert {\bf B} \vert \,\,\frac{\partial}{\partial z} \left(\frac{\vert {\bf B} \vert }{\rho} \right) > \rho^{\gamma} \frac{\partial}{\partial z} \left (\frac{T}{\rho^{\gamma-1}}\right)
\label{eq1}
\end{equation}

and

\begin{equation}
\vert {\bf B} \vert \frac{\partial \vert {\bf B} \vert}{\partial z} > \rho^{\gamma} \frac{\partial }{\partial z} \left( \frac{T}{\rho^{\gamma-1}} \right)
\label{eq2}
\end{equation}

The profile of the transverse field is chosen here so as to make the left-hand sides of Eq \ref{eq1} and \ref{eq2} positive in two different locations between $z=0.6$ and $z=0.8$. As a consequence, both the overlying atmospheric field and the layer below (possessing different field inclinations) will be unstable to buoyancy instabilities.

\begin{figure}[h!]
\centering
\includegraphics[width=7cm]{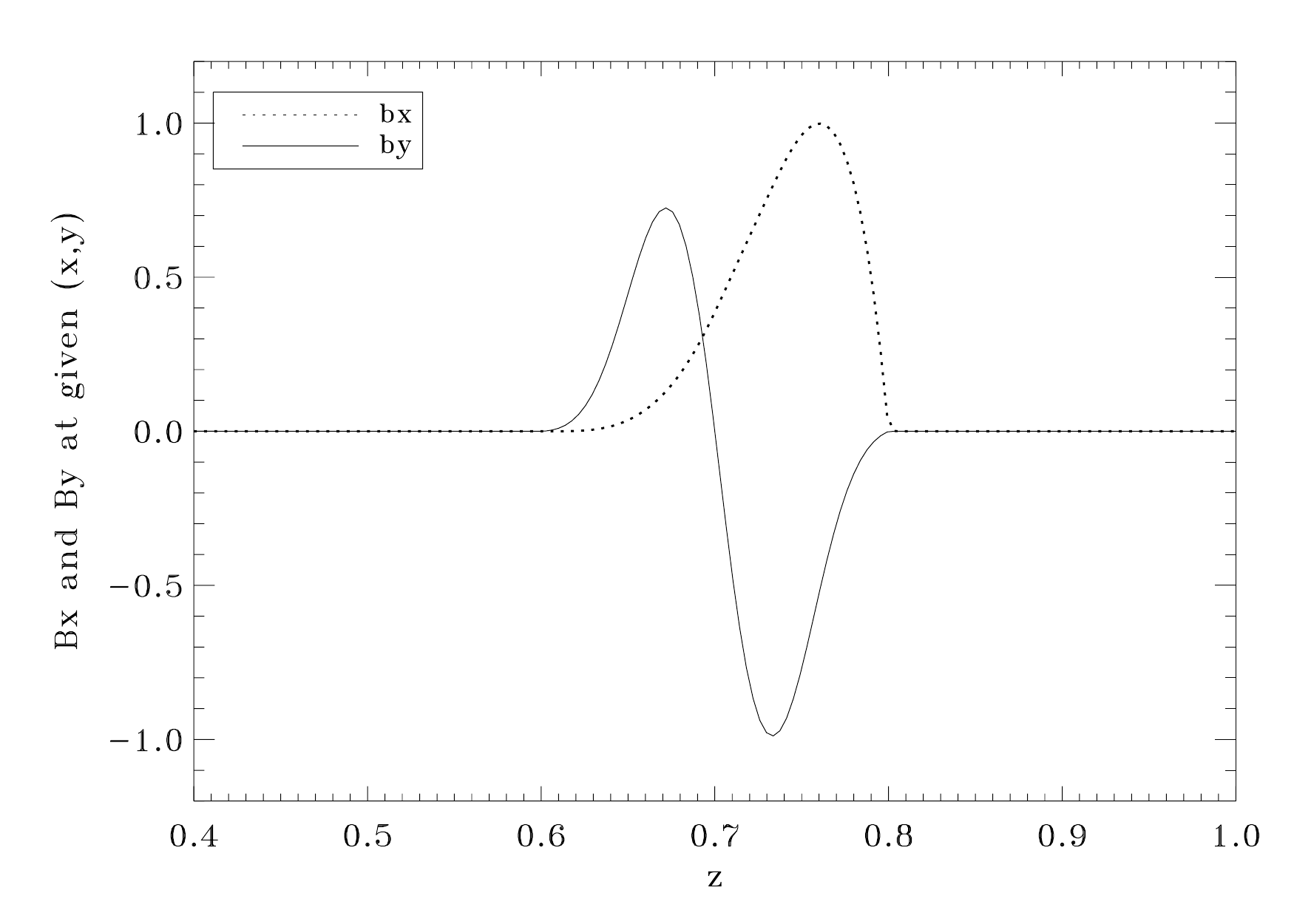}
\includegraphics[width=7cm]{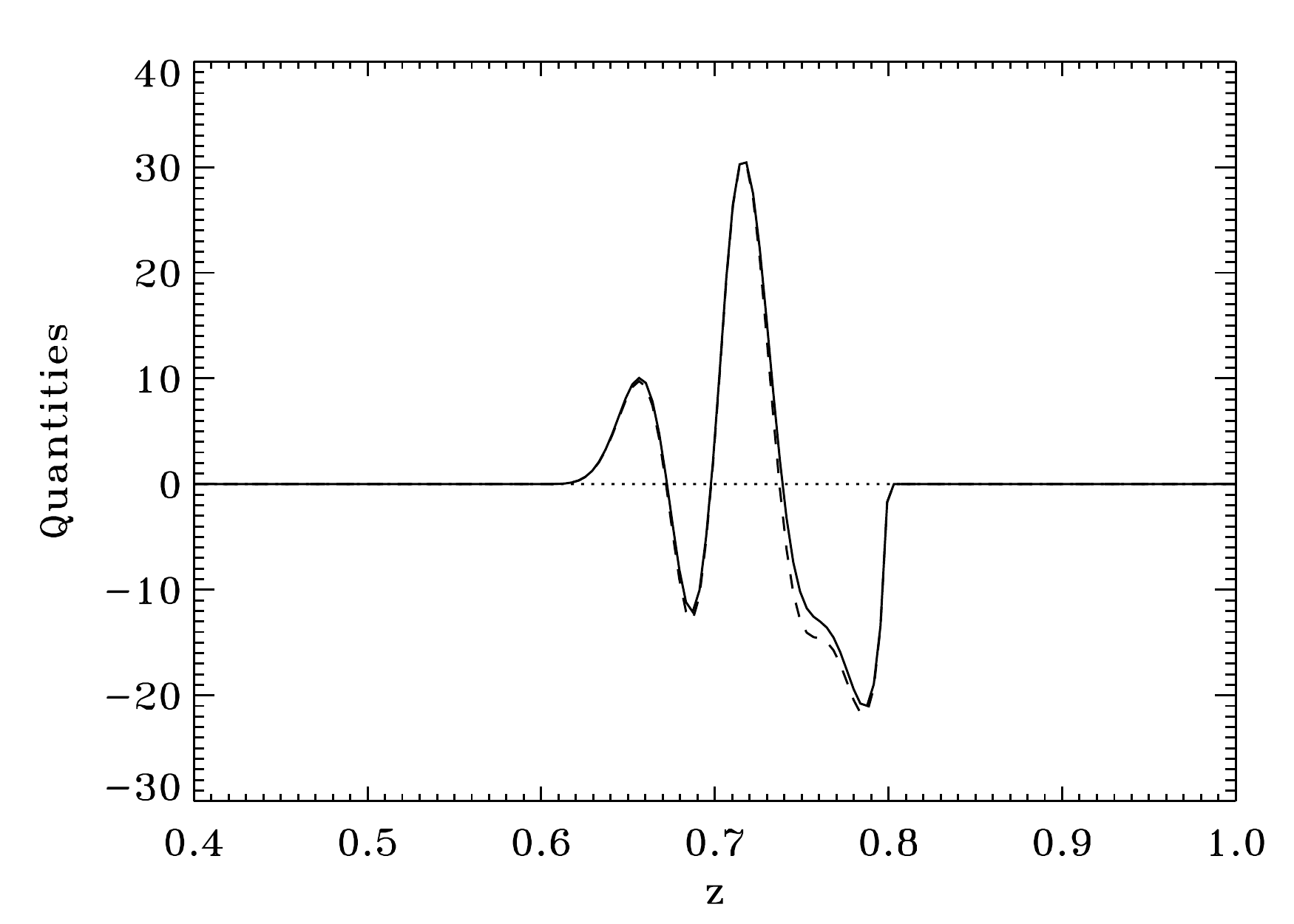}
\caption{Left panel: initial profiles of $B_x$ and $B_y$. Right panel: initial values of the right hand side and left-hand sides of Eqs \ref{eq1} and \ref{eq2}.}
\label{fig_1}
\end{figure}

As initial profiles for the magnetic field we choose the following:
$$
B_x=B_0 \, (z-0.6)^4\, (z-0.8)
$$
$$
B_{y}=B_{0y} \, (z-0.6)^{3/2}\, (z-0.8)\, \exp\left({-\frac{(z-0.7)^2}{0.05^2}}\right)
$$
Figure \ref{fig_1} shows the initial magnetic field profile as well as the characteristic quantities for the onset of magnetic buoyancy instabilities. It is clearly visible on the right panel of this figure that the magnetic field profile is made unstable at 2 different locations within the layer, without any interaction between them at the initial state. The profiles of $B_x$ and $B_y$ are smooth, the longitudinal field keeps the same sign within the whole layer while $B_y$ changes sign at $z=0.7$. Of course, in this case, the top unstable layer is dominated by a field in the $y$-direction, which would correspond to an almost entirely poloidal field at the base of the convection zone, and thus probably unlikely. Nevertheless, we are interested in the effect of a transverse field varying with depth and this configuration will thus capture the largest possible effects of this situation.

The dimensionless numbers used in this simulation are the following: $\gamma=5/3$ , $m=1.6$,$\theta=2$, $C_k=1.25 \times 10^{-2}$, $\sigma=2\times 10^{-2}$ for the hydro numbers and $F=1$, $\zeta=2\times10^{-3}$ for the magnetic quantities. The resolution used is $N_x=256$, $N_y=128$ and $N_z=260$.

\section{Results}

\subsection{Layer interaction}

\begin{figure}[h!]
\centering
\includegraphics[width=15cm]{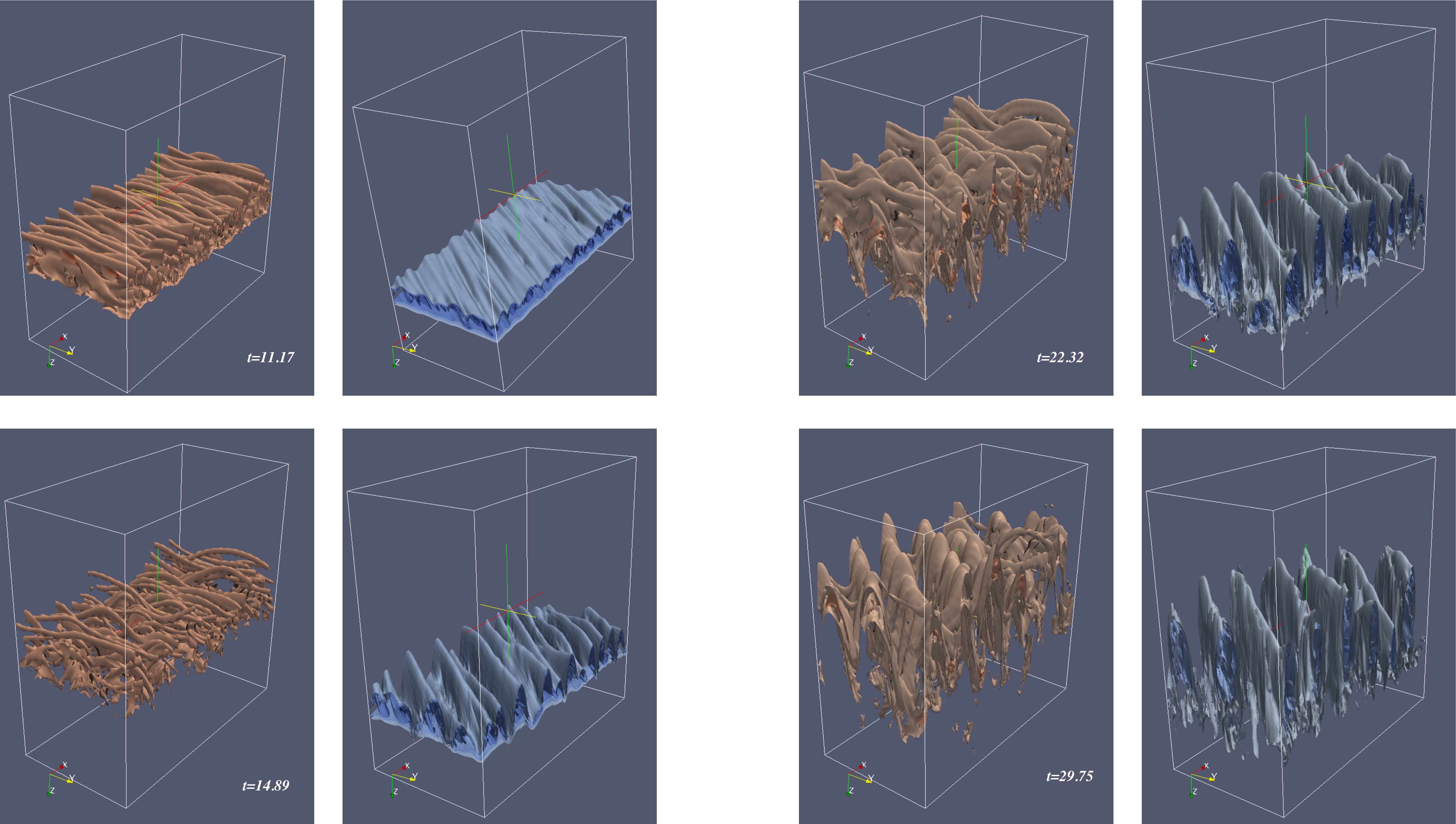}
\caption{Evolution of $B_y$ in the two unstable zones within the domain. $B_y$ is positive in the top layer (left panels) and negative in the bottom layer (right panels). The isosurfaces have dimensionless values between $-0.6$ and $0.6$, with a distance of $0.1$ between them. The colour scale is thus kept the same for all snapshots, allowing us to see the effect of magnetic diffusion on the rising layers. }
\label{fig_2}
\end{figure}

Figure \ref{fig_2} shows isosurfaces of the two unstable layers in this case. We focus on each unstable zone separately, in order to have a clearer idea of what the orientation of the field is in each layer and how the instability sets in. The two different orientations of the field appear clearly in this picture. We note that the positive $B_y$ zone (the top unstable layer) gets unstable before the other layer and very narrow flux tubes start to emerge. This is consistent with the linear analysis of \citet{Cattaneo90} and \citet{Favier12} which showed that the presence of an overlying atmosphere with inclined field lines reduces the growth rate of the instability of a magnetic layer. 

The two layers then interact and create 3D structures. At $t=14.89$, the structure of the top unstable layer is very interesting, small arched flux tubes are created and tend to maintain good coherence, a configuration which is quite favorable for the emergence of twisted structures. This is indeed what happens, as will be shown in the next subsection. Both layers now emerge with different speeds and rise towards the top of the layer. We find here that the top unstable layer rises on average more slowly than the bottom one. It is indeed affected more quickly both by viscosity and magnetic diffusion since the instability grows faster and since the flux ropes are smaller scale. As a consequence, at around 20-30 sound crossing times, the unstable modes of the bottom layer start to interact with those of the upper layer and the imprint of the preferred orientation of the modes of the blue layer can be seen on the red layer (top one). There is thus a strong interaction between the modes of various orientations and this will have consequences on the field line structure.

It is worth noticing that the whole layer gets completely disrupted by the magnetic buoyancy instabilities in this case, even though the resonant layer is located in the middle of the magnetic domain \citep[see discussion on the resonant layer by][]{Cattaneo90}. The vertical velocity (not shown here) indeed shows that downward motions are very strong, concentrated on narrow bands and penetrate very deep into the magnetic domain. It thus results in the complete disruption of the magnetized zone after a few tens of sound crossing times.

\begin{figure}[h!]
\centering
\includegraphics[width=15cm]{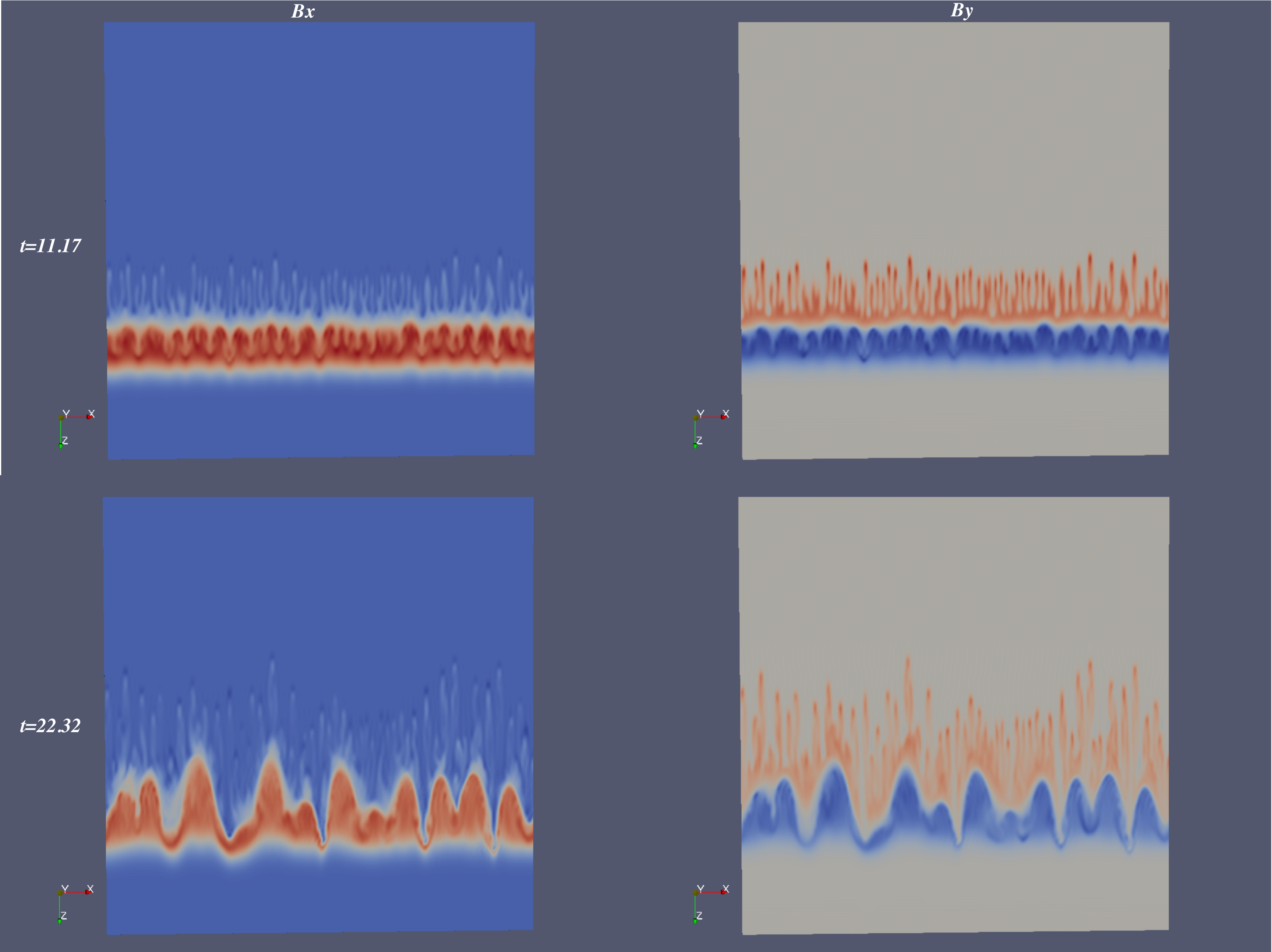}
\caption{Cuts at $y=cst$ of the magnetic field components $B_x$ and $B_y$. For $B_x$, the colour scale ranges between $0$ and $0.8$ and for $B_y$, the range is $-0.6/0.6$. Large-scale magnetic structures are clearly visible on both snapshots.}
\label{fig_3}
\end{figure}

Figure \ref{fig_3} shows cuts at $y=cst$ and at two specific times of the $x$ and $y$ components of the magnetic field. This picture shows the interaction between the very small scales emanating from the interchange instability of the top layer and the much larger scales below. As shown by \citet{Favier12} in their more idealized atmosphere, the magnetic tension lying at the interface between the two unstable layers promotes the merging between small-scale structures in the bottom magnetic layer. It is indeed quite clear that between $t=11.17$ and $t=22.32$, the top layer has been completely disrupted into small-scale features when the bottom layer has built much larger scales.

As a consequence, large-scale coherent structures are also generated in this less idealized case and could in principle rise through a large number of pressure scale heights before being diffused away or reprocessed by convective flows in a real stellar convection zone.

\subsection{Generation of twisted flux ropes}

\begin{figure}[h!]
\centering
\includegraphics[width=16cm]{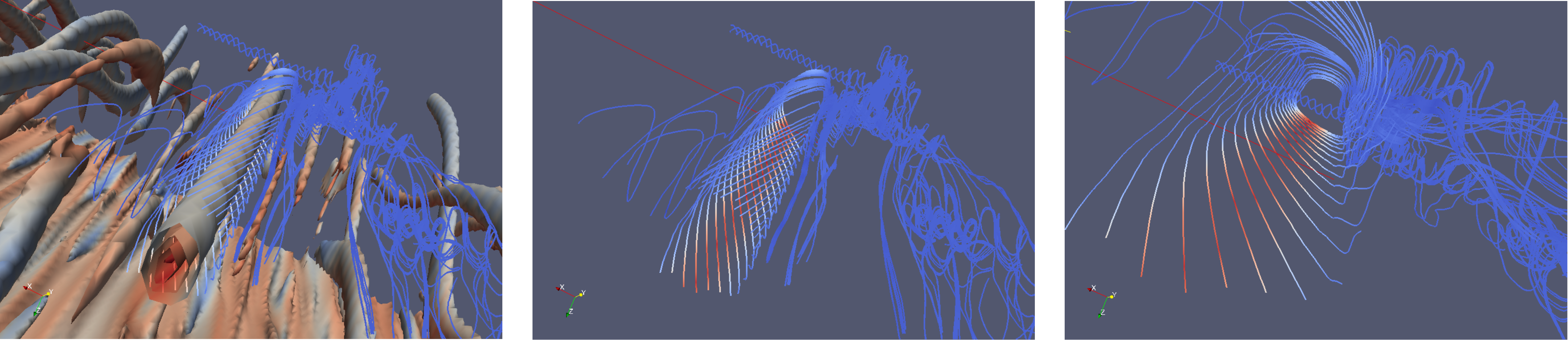}
\caption{Field line tracing around small flux ropes created in this simulation through the instability of a magnetically sheared layer.}
\label{fig_4}
\end{figure}

Finally, since our initial goal was to study the possible generation of twisted flux ropes in these calculations, we focus on the field line configurations, especially on the small unstable structures emerging in this simulation at the top of our magnetic domain. Figure \ref{fig_4} shows some rendering of magnetic field lines around current helicity concentrations in this calculation. As also shown by \citet{Favier12}, the initial configuration chosen here not only favors the generation of large-scale magnetic features from small-scale magnetic buoyancy instabilities but also promotes the creation of twisted flux ropes. Those helical fluxtubes should evolve in the polytropic atmosphere keeping much more coherence than the magnetic structures which are produced in a purely horizontal field case. Indeed, it was shown in multi-dimensional simulations of the dynamics of magnetic flux ropes \citep[e.g.][]{Emonet98} that a certain amount of twist of the field lines was necessary for concentrations of magnetic field to compete against the gravitational torque exerted on the rope during its emergence. Figure \ref{fig_4} shows that in these calculations, helical magnetic structures naturally emerge from the buoyancy instability in an initial configuration where both a longitudinal and a transverse field are present.

\section{Conclusion}

In this work, we studied the development of magnetic buoyancy instabilities of a magnetic layer in which two zones with different orientations of the field lines become unstable. The top layer becomes unstable first, in agreement with linear theory, and the two layers quickly interact, creating large-scale helical structures at the interface where magnetic tension and twist accumulate. The outcome of those calculations is similar to the case studied by \citet{Favier12} where a layer of field becomes unstable and interacts with a weakly magnetized (and thus stable) atmosphere. We thus show here that despite the more complex configuration of the atmosphere above (created here by the unstable top layer), twisted large-scale magnetic structures can still naturally emerge from buoyancy instabilities. This could in principle explain how strong concentrations of field created at the base of the solar (and stellar) convective zones are able to rise coherently to the surface to create the observed active regions.

\acknowledgements
We wish to thank S. Brun and B. Dintrans for the organization of Section 11 "Dynamo et magn\'etisme solaire et stellaire" of the SF2A 2012. We are also thankful to CALMIP for providing us with CPU time with which the calculations presented here were performed.


\end{document}